\documentstyle[11pt,aaspp4]{article}  %preprint format (single space)

\tighten
\slugcomment{To appear in
{\it The Astrophysical Journal}}

\begin{document}

%%%%%%%%%%%%%%%%%%%%%%%%%%%%%%%%%%%%%%%%%%%%%%%%%%%%%%%%%%%%%

%\magnification=\magstep1
%\hoffset=-0.5 truein
%\voffset=1 truein
%\vsize=8.5 truein
%\hsize=6.5 truein
%\parskip=0.0truecm
%\parindent=.25truein
%\raggedbottom
%\renewcommand{\baselinestretch}{2}
%\nopagenumbers
\baselineskip=24pt

\def\pmb#1{\setbox0=\hbox{#1}%
  \kern0.00em\copy0\kern-\wd0
  \kern0.03em\copy0\kern-\wd0
  \kern0.00em\raise.04em\copy0\kern-\wd0
  \kern0.03em\raise.04em\copy0\kern-\wd0\box0 }

\def\pp{\parshape=2 -0.25truein 6.75truein 0.5truein 6truein}

\def\ref #1;#2;#3;#4;#5{\par\pp #1 #2, #3, #4, #5}
\def\book #1;#2;#3{\par\pp #1 #2, #3}
\def\rep #1;#2;#3{\par\pp #1 #2, #3}

\def\undertext#1{$\underline{\smash{\hbox{#1}}}$}
\def\simlt{\lower.5ex\hbox{$\; \buildrel < \over \sim \;$}}
\def\simgt{\lower.5ex\hbox{$\; \buildrel > \over \sim \;$}}
\def\la{\lower.5ex\hbox{$\; \buildrel < \over \sim \;$}}
\def\ga{\lower.5ex\hbox{$\; \buildrel > \over \sim \;$}}
\def\r{\hangindent=1pc \noindent}
\def\j21{J_{-21}}
\def\lyal {{\rm Ly} \alpha}
\def\omigm {\Omega_{\scriptscriptstyle\rm IGM}}
\def\nhbr {N_{\scriptscriptstyle\rm HI,br}}
\def\nha {N_{\scriptscriptstyle\rm HI}}
\def\mj {M_{\scriptscriptstyle\rm J}}

\def\etal{{et~al.}}
\def\noi{\noindent}
\def\bs{\bigskip}
\def\ms{\medskip}
\def\ss{\smallskip}
\def\ob{\obeylines}
\def\l{\line}
\def\hrf{\hrulefill}
\def\hf{\hfil}
\def\q{\quad}
\def\qq{\qquad}
\renewcommand{\deg}{$^{\circ}$}
\newcommand{\um}{$\mu$m}
\newcommand{\uk}{$\mu$K}
\newcommand{\qrms}{$Q_{rms-PS}$}
\newcommand{\n}{$n$}
\newcommand{\cdmr}{${\bf c}_{\rm DMR}$}
\newcommand{\xrms}{$\otimes_{RMS}$}
\newcommand{\gt}{$>$}
\newcommand{\lt}{$<$}
\newcommand{\ldl}{$< \delta <$}

%%%%%%%%%%%%%%%%%%%%%%%%%%%%%%%%%%%%%%%%%%%%%%%%%%%%%%%%

%%%%%%%%%%%%%%%%%%%%%%%%%%%%%%%%%%%%%%%%%%%%%%%%%%%%%%%%
\bs
\bs
\bs
\bs
\bs
\bs
\bs
\bs
\title{Lyman Alpha Absorption Lines from Mini Pancakes}

\author{Biman B. Nath 
}
\noindent
\affil{Inter-University Centre for Astronomy \& Astrophysics,
Post Bag 4, Pune 411007, India}

\bs
\bs
\bs

\begin{abstract}

Recent numerical simulations show that many $\lyal$ absorption lines of 
column densities $\nha \la 10^{15}$ cm$^{-2}$ are produced in transient, 
mini pancakes. Such pancakes are modeled here, approximating the initial  
perturbation leading to the formation of the pancake as a single 
sinusoidal wave. The density and temperature profiles of the gas in 
the pancake are determined in the case where cooling and heating 
rate of the gas is
larger than the Hubble time, which is shown to hold for $z_c \sim 3$,
where $z_c$ is the collapse redshift.  
The $\lyal$ absorption line profiles for a line
of sight through the pancake are then calculated. The absorption lines in
general have wings signifying bulk motions in the gas. It is shown that
the deviation from a single Voigt profile is large for small H I column 
density lines, in which the effect of bulk motions is large. For lines
with $\nha > 10^{13}$ cm$^{-2}$, high temperature tend to 
wash out the signatures 
of bulk motion. 

The analytical modeling of mini pancakes associated with $\lyal$ forest
lines--- with $10^{13} \la \nha \la 10^{15}$ cm$^{-2}$---gives the 
corresponding mass scales. In structure formation models with CDM and with 
reheated IGM, the gas in the IGM is inhibited from falling into dark matter
potential wells with very small velocity dispersions. Structures with masses
larger than this limit (the Jeans mass) can have infall of gas and form mini 
pancakes. It is shown here that, for typical values of 
cosmological parameters, absorption lines with $\nha \sim 10^{14}$
cm$^{-2}$ correspond to structures with baryonic mass of $M_b \sim 10^{10}$
M$_{\odot}$ with an overdensity of $\sim 10$ at $z \sim 3$. The value of 
$\nha$ can change by a factor $\sim 3$ in the course of evolution of the 
pancake in time. It is also shown that there is an upper limit to
$\nha$ from a pancake due to the slow recombination rate and the importance of
collisional ionization at high temperatures. Mini pancakes do not give rise 
to $\lyal$ lines with $\nha \ga 10^{14.5}$ cm$^{-2}$, for a temperature of 
the IGM of $\bar T=10^4$ K, $\j21=1$ and $\Omega_{IGM} \sim 0.03$.

\end{abstract}


\section{INTRODUCTION}

Several ideas for the physical condition of the regions producing
QSO absorption lines with low H I column density (the $\lyal$ forest 
lines) have been proposed in the past. They have been, to name a few, 
pressure confined gas (Sargent \etal 1980; Ostriker and Ikeuchi 1983), 
gas behind intergalactic shock waves (Vishniac and Bust 1987),
mini halos (Rees 1986), caustics in velocity space (McGill 1990), and 
other scenarios. Recent numerical simulations have, however, shown that 
the physical condition is a mixture of many such scenarios (Cen \etal 1994, Hernquist \etal 1995, Miralda-Escud\`e \etal 1995, Zhang \etal 1995). The simulations, performed with a variety of structure formation models, show 
that transient, pancake-like regions of collapsing objects,--- and not
necessarily virialized objects---produce most of the $\lyal$ forest
lines with $\nha \la 10^{15}$ cm$^{-2}$.  For example,
Miralda-Escud\`e \etal (1995) find that the pancake associated with a 
typical $\lyal$ forest line has
a baryonic mass $\sim 10^{10}$ M$_{\odot}$ at $z \sim 3$, with
a thickness $\sim 50-100$ h$^{-1}$ kpc, and a transverse size
$\sim 1$ Mpc (where $H=100 \> h$ km s$^{-1}$ Mpc$^{-1}$ is the Hubble
constant at present epoch). The numerical simulations find excellent
fits with observations of the H I column density distribution of
$\lyal$ lines and the line profiles. (In the simulations, the Lyman limit 
and damped $\lyal$ systems in general come
from lines of sight through denser and more spherical objects.)

That such objects should be prevalent in the universe is not surprising.
`Zeldovich pancake' is the inevitable result of the fact that a near
simultaneous collapse of an object along all three directions, as in
a spherical collapse model, is very unlikely. Zeldovich (1970) pointed
out that a one dimensional collapse is more probable to occur. Zeldovich
and Sunyaev (1972) then considered the fate of gas inside such a nearly
two dimensional
pancake, although they considered very large masses ($\sim 10^{13}$ M
$_{\odot}$) which are the typical masses of the first collapsing
objects in a HDM universe. For a CDM like structure formation model,
the first objects to collapse have much smaller masses. Therefore,
in such a universe the existence of pancakes with smaller masses is not unexpected.

Such pancakes would have masses several times that of
the Jeans mass of the intergalactic medium (IGM). In a photoionized
IGM, with temperatures $\bar T \sim 10^4$ K, the sound velocity is $\sim
15$ km s$^{-1}$. The baryonic gas, therefore, does not fall into,
and cool to form galaxies in, dark matter potential wells with velocity
dispersions smaller than this sound velocity. Recent numerical
simulations support this scenario (Thoul and Weinberg 1995). 
The Jeans mass of the IGM, therefore, sets the mass scale above which
the baryonic gas can fall into the dark matter potential well.
The Jeans (total) mass is $\sim 5 \times 10^{10} \> 
h^{-1} \> (\bar T/10^4)^{1.5}$ M$_{\odot}$ at
$z \sim 3$ for such temperatures in the IGM in a $\Omega _o=1$ universe.
This corresponds to a Jeans length of $0.7 \, (\bar T/10^4)^{0.5} \, h^{-1} 
\, (1+z)^{-1.5}$ Mpc.
This is basically the length $\approx c_s \> (1/ (G \rho _{CDM})^{0.5}) $,
where the term inside brackets is the dynamical time of the halo
( see, e.g., eqn ($5$) of Reisenegger 
and Miralda-Escud\`e 1995). The (total) masses of the typical first 
collapsing objects at that redshift would, therefore, be several times 
this mass scale. It is no wonder then that the 
baryonic masses of the typical mini pancakes in the numerical simulations
are found to be of the order of $10^{10}$ M$_{\odot}$.

In CDM like theories, smaller structures will
also form, but due to the non-zero sound velocity of the baryonic gas,
the gas will not be clumped in such smaller structures. 

The idea of mini pancakes in the context of $\lyal$ lines
is also motivated by the recent observations
of coincident absorption lines in close quasar pairs. The observed
fraction (of $\sim \> 50-80 \%$) of coincident lines 
of two quasar pairs Q0107-025AB and Q1343+266AB, with proper separations
of $360 h^{-1}$ kpc and $40 h^{-1}$ kpc at redshifts $z \sim 1$ and
$z \sim 2$ (Dinshaw \etal 1994, Dinshaw \etal 1995, Bechtold \etal 1994),
show that the $\lyal$ absorbing systems have transverse sizes of the
order of a Mpc. With such large transverse sizes, the systems need
to have a small `thickness', {\it i.e.} the systems need to be flat,
so that the total amount of gas in $\lyal$ absorbers do not exceed
the limit on $\Omega_b$ from Big Bang nucleosynthesis (Rauch and
Haehnelt 1995).

Recently Rees (1995) has discussed the possible mass scales of the
pancakes associated
with $\lyal$ forest lines. He has suggested that mass scales between
$10^9$ and $10^{12}$ M$_{\odot}$, with various collapse factors
(in the range of $1-3$) are enough to cover the range of H I column
density of $\lyal$ clouds below those of Lyman limit systems. The present
work is aimed towards quantifying these speculations.

Zeldovich and Sunyaev (1972) discussed the behavior of gas in
pancakes approximating the 
initial perturbation that leads to the
formation of the pancake as a plane sinusoidal wave. They described
the density and temperature profiles of the gas as functions of time.
The goal of this work is to pursue the mini pancake model of the
$\lyal$ forest lines, along the lines followed by Zeldovich and Sunyaev
(1972) for large pancakes. Such an analytical picture will give us
the dependences of the various observables on different
parameters, which is sometimes difficult to understand from the results
of time consuming numerical simulations. The questions asked in this
paper relate to the physical conditions in these pancakes,
the relation between the mass scale of pancakes and their H I column density,
and also the evolution of the pancake gas in short time scales.

The plan of the paper is as follows. \S 2 discusses the physical
conditions of the pancake gas and determines the density and temperature
profile. \S 3 then calculates the $\lyal$ absorption line profile and
discusses the possible signatures of the pancakes in the observed
profiles. The different time scales for the pancake gas are then 
computed in \S 4, and some of the implications of these results
are discussed in \S 5.

\section{ GAS IN MINI PANCAKES}

\subsection{ Preliminaries}

If the mass of the collapsing object is larger than the current Jeans
mass, then the baryonic gas will also collapse with the dark matter.
In the Zeldovich approximation, the trajectory of a particle is given
by
\begin{equation}
{\bf r}=a(t) ({\bf q}+ b(t){\bf p} ( {\bf q})) \>.
\end{equation}
Here ${\bf r}(t, {\bf q})$ is the actual position of a particle as
a function of its Lagrangian (initial) coordinate ${\bf q}$ and time
$t$. The function ${\bf p} ( {\bf q})$ describes the initial perturbation,
$a(t){\bf q}$ is due to cosmological expansion, and $a(t)b(t)$ describes
the growth of the perturbation in the linear regime. It then follows that,
in the linear regime, 
the growth of the density perturbation in three spatial directions 
near a particle with a given ${\bf q}$ will be given by
\begin{eqnarray}
\rho ({\bf r}, t)&=& { \bar \rho (t) \over {\rm det} (\delta _{ij} +
b(t) (\partial p_j / \partial q_i))} \nonumber\\
& =&{ \bar \rho (t) \over \lbrack 1-b(t) \lambda _1({\bf q}) \rbrack \lbrack
1-b(t) 
\lambda _2({\bf q}) \rbrack \lbrack 1-b(t) \lambda _3({\bf q}) \rbrack
 } \> , \nonumber\\
\delta ({\bf r}, t) &\approx& b(t) (\lambda _1 + \lambda _2 + \lambda _3)
\>. 
\end{eqnarray}
Here, $\bar \rho (t)$ is the background density of the universe at
$t$, $\delta$ is the density
contrast, and $- \lambda _1, \, - \lambda _2, \,  -  \lambda _3$ are the 
eigenvalues of $(\partial p_j / \partial q_i)$. In the generic case,
the three eigenvalues will be different from one another. The
probability distribution of the eigenvalues was first derived by
Doroshkevich. The 
probability that $\lambda _1, \lambda _2, \lambda _3$ will all
simultaneously be either positive or all negative is $\sim 0.08$. The
probability that there is contraction in one direction and expansion
in other two, is $\sim 0.42$.

In the generic case, when the eigenvalues are different from one another,
the contraction will be strongest in the direction for which the 
corresponding eigenvalue is the largest among the three.
Eqns ($2$) show that, the density of the matter around the Lagrangian
coordinate ${\bf q}$ will diverge and the matter will be compressed
to a sheet in the (actual) Eulerian space, when one of the eigenvalues, 
say, $\lambda _1$, satisfies $b(t) \lambda _1 =1$. It can be shown that
the density profile in the case of one dimensional contraction is given
by $\rho \propto r^{-2/3}$ (see, e.g., eq. (2.7) in
Zeldovich and Shandarin 1989).

\subsection{ Gas infall and shock}

Although we use the Zeldovich-Sunyaev formalism for mini pancakes, we must
note the caveats involved. As mentioned above, mini pancakes are structures
with masses just above the Jeans scale and, therefore, gas pressure plays
an important role, unlike in the case of large pancakes discussed by
ZS72. However, for analytical simplicity, we neglect gas pressure in
this work.

For analytical simplicity, it is assumed that $\Omega =1$.
It is also assumed that the gas in the IGM is reionized (and reheated)
by the
time the mini pancakes form. The intensity of the UV background
radiation (at $912$ $\AA$) is written as $J=J_{-21} \, 10^{-21}$
erg s$^{-1}$ cm$^{-2}$ sr$^{-1}$ Hz$^{-1}$. The temperature, $\bar T$,
of the IGM gas roughly obeys $\bar T \propto (1+z)$ after reheating 
(Miralda-Escud\`e and Rees 1994).

The baryonic gas particles will initially follow the dark matter and
form a sheet. A shock wave will soon form which will travel outward. 
ZS72 (see also Zeldovich and Novikov 1983; Zeldovich and
Shandarin 1989) estimated the velocity of the
shock by assuming that the initial perturbation is
a plane sinusoidal wave. In reality, of course, the perturbation will
be more complicated. However, as shown below, the plane wave
approximation can be very useful in learning about the nature of
the gaseous pancakes. 

Pancakes are formed when there is a region in space 
where the velocity field is coherent.
Consider a pancake which is associated with such a region with 
a length scale of perturbation $L$,
and therefore, with some mass $M$ (the total mass initially contained in
the region of perturbation). The perturbation is assumed to be
a plane wave with a wavelength equal to $\lambda$. (Eqn (A2) in Appendix
A shows the relation between the mass and the wavelength.)
The derivation of the structure of the pancake using this approximation
by ZS72
is briefly described in the Appendix A below (after putting the necessary dependences on $h$ and the baryonic fraction
of matter $\Omega _b$;
note that their results are for a $\Omega _b =\Omega =1$ universe). 
Since the density of the IGM ($\Omega _I$) is more relevant for the gas 
inside the pancakes, than the general fraction of baryons in the
universe ($\Omega _b$), the former is used in the equations below.

Defining a parameter $\mu$ to be the fraction of the total mass $M$ 
that is inside the region bounded by the shock
layers, one can write the velocity with which the
gas runs into the matter in the pancake (in the laboratory frame)
(eqn (1) in ZS72; (A7) below), as,
\begin{eqnarray}
U &=&23.3 \> h^{1/3} \, (1+z_c)^{1/2} \, (\mu \pi )^{1/2} (\sin (\mu \pi))
^{1/2} \, \Bigl ( {M \over 10^{11} \, M_{\odot}} \Bigr )^{1/3} \> {\rm km/s}
\, ,
\nonumber\\
&\sim& 73.3 \> h^{1/3} \, (1+z_c)^{1/2} \,
\Bigl ( {M \over 10^{11} \, M_{\odot}} \Bigr )^{1/3} \, \mu \quad {\rm km/s}
\, , \qquad \mu \ll 1 \>. 
\end{eqnarray}
Here $z_c$ is the redshift of collapse of the perturbation. It can be
shown that (for $\Omega _o =1$, ZS72; (A5) below)
\begin{equation}
\Bigl ( {t \over t_c} \Bigr ) ^{1/3} = \Bigl ( {1+z_c \over 1+z} \Bigr )
^{1/2} = \Bigl ( { \sin (\mu \pi) \over \mu \pi } \Bigr )^{-1/2}
\>, 
\end{equation}
where $t_c$ corresponds to $z_c$. For $\mu \ll 1$, $\mu= {2 \over \pi}
((t/t_c) -1)^{1/2}$. The curve for eqn ($4$) and the above approximation
are plotted in Fig. 1 for $z_c=3$ and $z_c=4$. 
(It is shown below in \S 4 that the 
gravitational time scale of mini pancakes limit the possible values
of $\mu$ to be $\la 0.5$ .) Therefore, $\mu$ can be used as a
parameter equivalent to time $t$.

The velocity of the gas inside the shock front is assumed to be zero 
here, as it is expected to be small
(see Bond \etal 1984). The velocity of the shock front is then given
by, $V_{sh} = (1/3) U$ (Jones \etal 1981, Bond \etal 1984). Note that
$U \approx V_{sh}$ in ZS72.

For small values of $\mu$, the (particle) gas density in front of the
shock is (eqn (3) in ZS72; (A6) below),
\begin{eqnarray}
n _1 & \approx & 3 \, n _o \, {(1+z_c)^3 \over \pi ^2 \mu ^2} \, , \qquad
\quad \mu \ll 1 \, ,
 \nonumber\\
& = & 2.3 \times 10^ {-7} \, h^{-2} \,(1+z_c)^3 \, \mu ^{-2} \, 
\Bigl ({\Omega _I \over 0.01} \Bigr ) \>, 
\end{eqnarray}
where it is assumed that the IGM gas is ionized ($\rho = 0.6 \, n \, m_p$)
and $n_o$ is the particle density at present epoch. The postshock
gas density is $n_{sh}=4n_1$, for a strong shock. The postshock
temperature is, again for small $\mu$,
\begin{equation}
T_{sh}= {0.6 m_p U^2 \over 3 \, k} = 1.3 \times 10^5 \, h^{2/3} \, (1+z_c)
\, \Bigl ( {M \over 10^{11} \, M_{\odot}} \Bigr ) ^{2/3} \, \mu ^2 \> 
{\rm K} \>. 
\end{equation}
The pressure behind the shock front ($p_{sh}= (4/3) \, \rho _1 \, 
U ^2$) is 
a independent of $\mu$ (i.e., of how much of matter is inside the
shocked layer for a given pancake ) for small $\mu$ (ZS72).

The distance of the shock front from the central plane of the pancake
can be estimated as,
\begin{eqnarray}
r_{sh}&=&\int V_{sh} \, dt = \int V_{sh} \, {dt \over d \mu} \, d \mu
\nonumber\\
&\approx & 0.27 \, h^{-2/3} \, (1+z_c)^{-1} \, \Bigl ( {M \over 10^{11} 
\, M_{\odot}} \Bigr ) ^{1/3} \, \mu^3 \> {\rm Mpc} \, , 
\qquad\quad \mu \ll 1 \, .
\end{eqnarray}

As mentioned above and explained in the Appendix, the above expressions
are obtained in the absence of gas pressure. As gas pressure is more
important for mini pancakes than the large pancakes studied by ZS72,
these expressions are only very approximate.

\subsection{ Density and temperature profile}

What is the maximum gas density in the pancake? The maximum density would
be infinite for infinite cooling rate and for a zero pressure gas outside,
neither of which is true in the present case. It can be shown that the
cooling time scale of the gas
%, due to recombination ( the cooling 
%due to line emissions is suppressed in the presence of the UV background %radiation), 
in the presence of the UV background
is larger than the Hubble time  at $z \sim 3$ for the relevant gas 
densities of this problem (see Appendix B). 
%For $\j21=1$ and $n \la
%10^{-4}$ cm$^{-3}$ the cooling rate is dominated by recombination.
%Using the recombination cooling rate given in Black (1981), 
%it follows that at $z_c =3$,
%the cooling time scale is shorter than the Hubble time only if
%$(n/ 5 \times 10^{-4}) \, (T/5 \times 10^4)^{-0.7} \, h^{-1} \, > 1$. 
%This condition is not satisfied for typical pancakes (see below and in \S 4).

%In the absence of any appreciable cooling, the gas, therefore, contracts
%adiabatically. 

However, the heating time scale becomes comparable to or shorter than
the Hubble time for $z_c \sim 3$ (Appendix B). In the following, we restrict
our analysis to $z_c \sim 3$ where both cooling and heating time scales are
larger than the Hubble time and assume that the gas contracts adiabatically.

Furthermore, since the evolution of the IGM temperature $\bar T$ depends
on various parameters and models of ionization (Miralda-Escud\`e and Rees
1994), we will, for simplicity, assume a
constant IGM temperature of $\bar T \approx 10^4$ K in the following.
Introducing an evolution of $\bar T$ will introduce more free parameters,
which we will avoid.

 If the ambient
IGM density is $\bar n$, with pressure $\bar p$ (and temperature $\bar T$), 
then the maximum
gas density and temperature in the pancake is given by
\begin{eqnarray}
n_{max}&= & \bar n \Bigl ({p_{sh} \over \bar p} \Bigr )^{3/5}
=1.2 \times 10^{-4} \, h^{12/5} \, \Bigl( {1+z_c \over 4} \Bigr )^{18/5}
\, \Bigl ( {M \over 10^{11} \, M_{\odot}} \Bigr ) ^{2/5} \, \Bigl (
{\Omega _I \over 0.01} \Bigr ) \> 
\Bigl ( { \bar T \over 10^4 \, {\rm K}} \Bigr )^{-3/5}, \nonumber\\
T_{max}&=&\bar T \Bigl ({n_{max} \over \bar n} \Bigr )^{2/3}
=4.7 \times 10^4 \, h^{4/15} \, \Bigl( {1+z_c \over 4} \Bigr )^{2/5}
\, \Bigl ( {M \over 10^{11} \, M_{\odot}} \Bigr ) ^{4/15} \, 
\Bigl ( { \bar T \over 10^4 \, {\rm K}} \Bigr )^{3/5} 
\end{eqnarray}

As mentioned in \S 2.1, in the Zeldovich approximation, the density  
of the gas has a profile (in the direction perpendicular to
the pancake) that is given by $\rho \propto r^{-2/3}$
(e.g., eqn. (2.7) of Zeldovich and Shandarin 1989).
The density profile of the gas
can, therefore, be divided into four characteristic regions: 
(a) in front of the shock,
the density follows a $r^{-2/3}$ profile, as explained above; 
(b) the density jumps by a factor of $4$ at the shock front; 
(c) behind the shock, the gas
again follows a $r^{-2/3}$ profile, in the Zeldovich approximation,
as explained above;
(d) and then it reaches a maximum when the density is $n_{max}$ of eqn ($8$).
 For analytical simplicity, we assume this central
region, say, $r < r_o$, with $n \sim n_{max}$ to be a region of 
uniform density. 

In the single wave approximation used here, we have neglected the motion
of the gas in the directions along the pancake. In reality, gas will
also move along the pancake. 
Modelling this effect is, however, beyond the scope of the single wave approximation. We
neglect it here, and assume that the density in the central region 
is a constant in time, for the short time evolution we are interested
in here ($\mu \ll 1$). Gas will also tend to expand in the perpendicular
direction, but for small $\mu$ (which is considered in the present work), 
pressure behind the shock$p_{sh}$, stays constant with $\mu$
(that is, in time) and the ram pressure will tend to confine the gas.
In other words, we assume that,
\begin{eqnarray}
n & \sim & n_{max}  \qquad\qquad\qquad\quad r < r_o \nonumber\\
& \propto & n_{max} \, ({r \over r_o})^{-2/3} \qquad\quad
 r_o < r < r_{sh} \nonumber\\
&\sim & n_{sh} \qquad\qquad\qquad\qquad r \sim r_{sh} \nonumber\\
&\propto &{n_{sh} \over 4} \, ({r \over r_{sh}})^{-2/3} 
\quad\qquad r> r_{sh} \>.
\end{eqnarray}
If $r_{sh} < r_o$, then $n \sim n_{max}$, $r <r_{sh}$, and
$n \propto {n_{sh} \over 4} \, ({r \over r_{sh}})^{-2/3}$ for $r > r_{sh}$.
Matching the densities at the various boundaries,
as explained above, one obtains for the extent of this central
region, as
\begin{eqnarray}
r_o= 3 \, h^{-19/15} \, \Bigl( {1+z_c \over 4} \Bigr )^{-19/10}
\, \Bigl ( {M \over 10^{11} \, M_{\odot}} \Bigr ) ^{-4/15}
\> \> {\rm kpc} \>.
\end{eqnarray}
This length scale will be important in calculating the column densities
in \S 2.4.
%In reality, cooling of the gas will somewhat steepen the profile
%in the central region, but such effects are neglected here considering
%the slow cooling rate.

Similarly, the temperature has a maximum value in the central region,
as given in eqn ($9$). At the current position of the shock, $r_{sh}$,
the temperature will be given by eqn ($6$). For $r_o < r < r_{sh}$, 
the gas would have been shocked and
would be raised to corresponding $T_{sh}$. As $T_{sh} \propto \mu ^2$
and $r_{sh} \propto \mu ^3$, it is easy to see that $T \propto
r^ {2/3}$ for $r_o < r < r_{sh}$. Outside $r_{sh}$, the temperature
the same as that of the IGM ($\bar T$). 
We remark here that the temperature profile will in general change in time
due to cooling.
As explained above, the cooling rate is small for $z \ga 3$ and the
above temperature and density profiles hold only in that regime.
%Gas within
%$r<r_o$, therefore, suffers only adiabatic contraction, and only
%when $r_{sh} \ga r_o$, the still infalling gas goes through a
%shock front.

%At an instant of time, the gas pressure is not constant within the
%shocked region ($p \propto n^{5/3} $). 
%In contrast, the large pancakes studied in the context of HDM models,
%had hotter shocked regions and larger sound velocities, and therefore with
%a near constant pressure (Bond \etal 1984).

The profiles of $\delta =n /\bar n$, $\delta _{HI} = n_{HI} / \bar
n_{HI}$ and $T / \bar T$  for a typical pancake are shown in Fig. 2,
for $\mu=0.3, 0.4$. We have used $M=5 \times 10^{11}$ M$_{\odot}$, $h=0.75$,
$z_c=3$ and $\Omega _I=0.03$. For each value of $\mu$, the values of
$r_o$ and $r_{sh}$ are first evaluated (eqns ($7$) and ($10$)), 
and then the density at each point in space
using eqns ($5$) and ($8$). The density of neutral hydrogen is evaluated
at each point of the pancake using ionization balance (see \S 2.4).
The increase in the overdensity at later times
is due to the decrease in the ambient density in the universe
in time. It is seen that the gas in the pancakes have a typical
maximum overdensity of order $\sim 10 \hbox{--} 50$ 
(see also Miralda-Escud\`e \etal 1995). One should note that
in reality gas will expand along the pancake, thereby decreasing the central
density. Therefore, the overdensity obtained above should be taken as 
an upper limit.

The density and temperature profiles in Fig. 2
are much different from those in large, massive pancakes, as studied
by ZS72, in that the maximum 
density in their case is much larger, and therefore, $r_o$ is smaller
(it was small enough to be unimportant for those pancakes).
Also, $T_{max}$ in ZS72 was much lower because a cold IGM gas was assumed. 
For small $\mu$, the temperature profile does not have 
the usual `double
horn' profile that is expected in large pancakes (Jones, Palmer and Wyse
1981). Such a profile is possible if, $
\mu \ga 0.3 \, h^{-3/15} \, ((1+z_c)/4) ^{7/10} \, (M/ 10^{11} \, M_{\odot})
^{3/15} \, (\bar T / 10^4 \, K)^{1/2}$. As discussed below, large
values of $\mu$ in general correspond to large $\nha$.
Therefore, such temperature profiles are not expected for small H I column
density $\lyal$
lines. The spectra and the corresponding density and temperature profiles
obtained in the numerical simulations of Miralda-Escud\`e \etal (1995)
support this.

The extent, or the thickness, of the pancake can be estimated by
finding the distance $r'$ where the density contrast $\delta 
=(n- \bar n) / \bar n$ (where $\bar n$ is the background IGM density)
is unity. For the above profile, this happens at $r' \sim 
({2 \pi ^2 \mu ^2 \over 3})^{-3/2} \> r_{sh} \sim 6.2 \times 10^{-2} \,
\mu ^{-3} \, r_{sh}$. For $M=3 \times 10^{11}$ M$_{\odot}$
(corresponding to $M_b \sim 10^{10}$ M$_{\odot}$, for $\Omega _I 
\sim 0.03$), $h=0.5$ and $z_c=3$, the `thickness' of the pancake
 is $\sim 20$ kpc. For comparison, the typical thickness for pancakes as
reported by Miralda-Escud\`e \etal (1995) is $50-100$ $h^{-1}$ kpc. 
It should be noted that
this is also consistent with the upper limit on the thickness
($\sim 50$ kpc) put from the combined knowledge of the 
 large transverse sizes of
$\lyal$ absorption systems and the value of $\Omega _b$ from
Big Bang nucleosynthesis (Haehnelt and Rauch 1995).

\subsection{Neutral and H I column density}

For a line of sight which is perpendicular to the pancake, the
total column density of the gas is,
\begin{eqnarray}
N_H & \sim &2 \, \int _{r=0} ^{r=r_{sh}} n(r) \, dr \, = 2 \,
n_{max} \, r_o \, \lbrace 1 + 8.8 \, h^{1/5} \, \Bigl( {1+z_c 
\over 4} \Bigr )^{3/10} \, \Bigl ( {M \over 10^{11} \, M_{\odot}} 
\Bigr ) ^{1/5} \, \mu \rbrace \nonumber\\
& \approx & 2.3 \times 10^{18} \, \Bigl ( {\mu \over 0.3} \Bigr ) \,
 \Bigl ({h \over 0.5} \Bigr )^{4/3} \, \Bigl( {1+z_c 
\over 4} \Bigr ) ^2 \, \Bigl ( {M \over 10^{11} \, M_{\odot}} 
\Bigr )^{1/3} \, \Bigl ({\Omega _I \over 0.01} \Bigr ) \> \>
{\rm cm}^{-2} \>. 
\end{eqnarray}
This is comparable to the values of column density plotted in Fig. 9a of 
Miralda-Escud\`e \etal (1995). For a gas in ionization equilibrium
with photoionization by the UV background radiation, and with 
collisional ionization,
one can also estimate the H I column density. Photoionization and
recombination rates are taken from Black (1981), and Cen 
(1992). Fig. 3 (a) and (b) 
 plot $\nha$ and $N_H$ vs. the mass (total and baryonic) that corresponds 
to the pancake, for a line of sight perpendicular to the pancake. Here
$M_b=\mu \> \Omega _I \> M$.
Fig 3 (c) plots $N_H$ against $\nha$. Fig. 3 (d) shows the evolution
of $\nha$ in time, or equivalently, with increasing values of $\mu$.

The plots show that, in general, larger (more massive) pancakes
correspond to higher $\nha$ (approximately, $\nha \propto M^{2/3}$
(see below in \S 4 for a discussion on this). There is however a scatter,
due to possible differences in the ambient density ($\Omega _I$),
collapsing redshift ($z_c$),
and due to the evolution of pancakes in time. Since $N_H \propto
\Omega _I$, for a photoionized gas, $\nha \propto \Omega _I ^2$
(see below). The column density
increases at higher collapse redshifts, but this can be an artifact
of the assumption that a constant background temperature ($\bar T$).
In reality, $\bar T$ may increase with redshift (in a photoionized
universe $\bar T \propto (1+z)$ much after the reheating epoch). This
will decrease the density in the central region and will decrease
the column density. Such an evolution in the IGM temperature introduces
another parameter for the pancakes, and will not be discussed here
to avoid confusion.
The evolution of the individual pancakes first increases $\nha$ till 
about $\mu \sim 3$
(for $M \ga 10^{11}$ M$_{\odot}$),
after which hotter temperatures deplete more neutral atoms and halts the
increase in $\nha$. 

This behavior is also reflected in Fig. 3(c) which plots $N_H$ against
$\nha$. Large values of $\nha$ are either associated with large $M$
or large $\mu$, both of which are associated with high temperature,
which tends to decrease the neutral fraction. This fact, therefore, puts
an upper limit on the value of $\nha$ for $\lyal$ lines from pancakes.
In particular, for $\j21=1$ and $\Omega_I \sim 0.03$, mini pancakes 
can only give
rise to $\lyal$ lines with $\nha \la 10^{14}$ cm$^{-2}$ for $z_c=3$. For 
$z_c=4$, this limiting value is found to be $\nha \la 10^{14.5}$ cm$^{-2}$.

%Apart from a small scatter in the plot of $N_H$ vs. $\nha$,
%approximately, $N_H \propto \nha ^{0.5}$, for small $\mu$, and $N_H
%\propto \nha$ for large $\mu$. This is because of the fact
%that for a photoionized gas
%$n_H \propto n^2 T^{-0.7}$, and initially, when the gas is simply
%adiabatically compressed and not shocked, the temperature of the warm gas
%remains almost a constant. This explains the behavior ($N_H \propto \nha
%^{0.5}$) at small $\mu$. It is only later, when the shock appears,
%that the temperature changes with time, and, therefore, with 
%$\nha$, thereby changing
%the slope.  
%This change in slope due to collisional ionization introduces a maximum 
%value of $\nha$  as seen in Fig. 3(c).  

%%%%%%%%%%%%%%%%%%%%%%%%%%%%%%%%%%%%%%%%%%%%%%%%%%%%%%%%%%%%%
%%% Section 3 %%%%%%%%%%%%%%%%%%%%%%%%%%%%%%%%%%%%%%%%%%%%%%%
%%%%%%%%%%%%%%%%%%%%%%%%%%%%%%%%%%%%%%%%%%%%%%%%%%%%%%%%%%%%%
\section{ ABSORPTION LINES}

Armed with the knowledge of the density and temperature profiles, one
can calculate the $\lyal$ absorption line profile for a line of
sight through the pancake. The additional information necessary
is that of the velocity field. The bulk velocity of the gas within
the shocked layers is small and is neglected here. Outside the shocked
layer, the gas particles  stream towards the pancake with
velocity that is determined by the initial velocity perturbation. In
Zeldovich approximation, the velocity of a particle remains constant.
This allows one to find the velocity profile of the gas at later
times. 

Consider a particle initially at Lagrangian (comoving) coordinate
$\xi$ with velocity $v$. This particle will eventually meet the shock
front with the same velocity $v$ at some point $r$ in the Eulerian 
(physical) coordinate. This velocity $v$ and the point $r$ are
given by eqns ($3$) and ($7$). 
Also, as described in Appendix, every point
$\xi$ is associated with a value of $\mu$. Since the velocity remains 
constant, 
this mapping is also true for other points in physical space. For any
other point $r'$, one needs to find the corresponding $\xi '$
or $\mu '$ from eqn ($7$) and then
find the velocity from eqn ($3$).

Let $r=0$ be the origin (in the dimension perpendicular to the
pancake) at some redshift $z$. The total velocity of a particle at 
$r$ is therefore, $V(r)=Hr+v(r)$ where $H$ is the Hubble
constant at $z$ and $v(r)$ is the peculiar velocity. The profile
of the absorption line is determined by the overdensity of H I
in velocity space, convolved with a Gaussian of width $b(r)
=\lbrace 2 k T(r)/ m_p \rbrace ^{0.5}$. The smoothed H I
overdensity at some velocity $V'$ in velocity space is given by
(Miralda-Escud\`e and Rees 1992),
\begin{equation}
\delta _{\scriptscriptstyle\rm HI} (V')= \int \, dr \, \delta
_{\scriptscriptstyle\rm HI}(r) \, H \, \exp \Bigl ( -
{(V(r)-V')^2 \over b(r)^2 } \Bigr ) \, (\pi ^{1/2} \, b(r))
\>. 
\end{equation}
If the Gunn-Peterson optical depth to $\lyal$ absorption by the
IGM is $\tau _{GP}$, then the total optical depth at velocity
$V'$ is given by $\tau (V')= \tau _{GP} (1+ \delta 
_{\scriptscriptstyle\rm HI} (V'))$. Here, $\tau _{GP}$ is the
Gunn-Peterson opacity, and in a photoionized universe it is
given by $\tau_{GP}=0.32 \> \Omega _I ^2 h^3 \> (1+z)^{4.5} \> \j21
^{-1}$ ( using eqn ($39$) of Miralda-Escud\`e and Ostriker (1990),
with clumping factor $f=1$ and the spectral index $\alpha =1$).

Examples of line profiles are plotted in Fig. 4, for various
values of $\mu$ for a particular pancake, and for different
orientations. Also shown are the fits to the profiles using
Voigt function, which depend on $\nha$ and a $b$ parameter.
 The value of $\nha$ 
 changes when the angle of inclination $(\theta)$of the 
pancake to the line of
sight is changed, and is roughly proportional to $(\cos (\theta))$.
(The total H~I column density is exactly proportional to  $(\cos (\theta))$,
but this is only approximately true for the H~I column density of the fitted central line, as some of the neutral atoms also show up in the extra lines.) This is mainly because of our assumed flat density profile for $r \la r_o$.

It is seen that the absorption lines from pancake can be fit by Voigt
profiles, although with some deviations. In general, the
absorption profile consists of a central dip, which follows a Voigt
profile, and some extra dips which are shifted from the central one
in frequency space. These wings make the value of $\nha$ from fitting
the central portion of the profile smaller than the corresponding value
from Fig. 3 (d). Absorptions by some neutral atoms
 show up in the profiles away from the central portion -- causing the
absorption profile to deviate from the Voigt profile.

Even though pancakes are associated with bulk motions, 
the thermal gas makes the profile close to Voigt profile for lines
with large $\nha$.
Lines with small $\nha$, however, are in general associated with
small values of $\mu$, when only a small fraction of the total
mass has been heated through adiabatic compression. One,
therefore, expects to see the deviations from a simple Voigt profile
in the case of low mass pancakes, or when $\mu$ is small in a pancake,
when bulk motion dominates over the thermal contribution to $b$.
Such an example is shown in Fig. 4(d). It is similar to some of the
low column density lines in the spectra obtained in the numerical
simulations of Miralda-Escud\`e \etal (1995) (see their discussion
in \S 3.3.3). 

 The deviation of the absorption profile from
a Voigt profile can be quantified by the ratio of the equivalent width
of the central dip with the Voigt profile ($W_c$) to the total equivalent
width in the absorption profile ($W_{tot}$). 
As shown in Fig. 5a, this ratio is very small for lines with small
total equivalent widths. These lines are produced by
less massive pancakes and for pancakes with small
values of $\mu$ (and therefore with smaller values of $\nha$). The
curves show that a fraction of $\ga 0.1$ of the total equivalent width
of the absorption line shows up in the wings. 
The wings could, however, be fitted separately as extra lines
with Voigt profiles, but the interpretation of these (small $\nha$)
lines as separate `objects' or `clouds' would not be appropriate, as these
extra lines simply reflect the bulk motion in the pancake
(see also Rauch 1995). For $M \ga 5 \times 10^{11}$ M$_{\odot}$, the
ratio $W_c/W_{tot} \la 0.5$ for $W_{tot} \la 0.1$ $\AA$, corresponding
to $\nha \la 2 \times 10^{13}$ cm$^{-2}$. Several
of the observed small $\nha$ ($\nha \la 10^{13}$ cm$^{-2}$) lines,
therefore, reflect the bulk motions relevant for the formation of
mini pancakes (see also McGill 1990, where it was pointed out that
velocity cusps in redshift space can masquerade as $\lyal$ lines).

For small deviations from the Voigt profile, the $b$ value of the
central line is determined by thermal motion of the atoms. Fig 5b
plots the ratio of the $b$ from Voigt profile fitting of the central
line, to the $b$ that corresponds to $T_{max}$ of eqn ($6$), as 
a function of $\nha$ of the central line (determined from profile
fitting). 
It clearly shows that for lines with $\nha \ga 10^{13}$ cm$^{-2}$,
the b value is determined by the thermal motion of the atoms.
For lines with $\nha \ga 10^{14}$ cm$^{-2}$, however, the temperature in
the shock front dictates the value of $b$ (and therefore deviates
from the value of $b$ corresponding to $T_{max}$ of the central region).
A statistical study for the distribution of b is, however, beyond
the scope of the present work.

%%%%%%%%%%%%%%%%%%%%%%%%%%%%%%%%%%%%%%%%%%%%%%%%%%%%%%%%%%%%%
%%% Section 4 %%%%%%%%%%%%%%%%%%%%%%%%%%%%%%%%%%%%%%%%%%%%%%%
%%%%%%%%%%%%%%%%%%%%%%%%%%%%%%%%%%%%%%%%%%%%%%%%%%%%%%%%%%%%%
\section{ DISCUSSIONS}

%Figs. 2 (a) and (b) show that, approximately, $\nha \propto M^{2/3}$. One
%can understand this qualitatively in the following way. Consider 
%pancakes of constant width $w$ (extent in the direction of the line of 
%sight). If the gas has not moved substantially in the transverse
%direction, then the transverse size $L \propto M^{1/3}$, since
%$M$ is the mass contained in the region of perturbation of length
%scale $L$. Then, the total mass, $M \, \sim \, n \, w \, L^2$, which
%implies that, $n \, \propto \, M^{1/3}$. The H I column density, therefore,
%$\nha \, \sim \, n_H \, w \, \propto n^2 \, w \, \propto \, L^2
%\, \propto \, M^{2/3}$.

In CDM type scenarios the mass scale of the first gaseous pancakes
is given by the Jeans mass. The cutoff in the power spectrum, however, is 
not
as sharp as the lower cutoff in HDM scenarios. Also, the power spectrum
in the vicinity of the Jeans cutoff is flatter for CDM type structure 
formation models than the HDM power spectrum. This means that the masses
of the first gaseous pancakes may not have a single value and is likely
to have a distribution. 

It is not clear whether or not the column density distribution of $\lyal$
lines arises from this distribution in mass. If that is the case, 
then the above relation can be used, in principle, to calculate the 
distribution of $\nha$ (number of lines with H I column density
$\nha$ per unit H I column density per unit redshift, or ($d N/d \nha \, 
dz$)),
if an expression for the distribution function of the pancakes can
be found in terms of their masses (or, equivalently, the length
scale $L$ of the region in which the initial velocity field is coherent).
However, the cause of the column density distribution of $\lyal$
lines is likely to be more complicated than this. Firstly, if
the pancakes survive for about a Hubble time, the evolution of the
pancakes need to be taken into account.
In any case, such a statistical study is beyond the scope of this
work.

 Several observations involving double lines of sight suggest that
the sizes of regions in the direction perpendicular to the line of sight
can be several hundreds of kpc. Miralda-Escud\`e \etal (1995) find the
typical transverse size of pancakes to be $\sim 1$ Mpc. As mentioned earlier, the Jeans scale at $z \sim 3$
is $\lambda _J \sim 0.13 \, (\bar T / 10^4 \, K)^{0.5} \,
 (h / 0.75)^{-1} $ $\, ((1+z)/ 4)
^{-1.5}$ Mpc, for a photoionized universe with $\bar T \propto (1+z)$.
The typical first pancakes with infalling baryonic gas will therefore
have sizes several times this (the diameter will be twice this value).
It is, therefore, conceivable that the transverse size of pancakes
would be $\sim 1$ Mpc. More over, the gas flows in the transverse direction
will make the size even larger. As mentioned earlier, when one of the
eigenvalues ($\lambda _1, \,  \lambda _2, \, \lambda _3$) is negative
(which describes contraction), the probability that the other two eigenvalues
are positive (that is gas {\it expands} in the transverse direction), is
$\sim 0.42$ (ZS72). As suggested by many previous authors, this expansion
can make the pancakes larger in the transverse direction, and increase the
probability of detection. 

Finally, a word about the approximation used is in order. The results
obtained here should be considered as only very approximate, and
have large error bars, because of the approximation of single plane
waves. However, it is interesting that some information about the
behaviour of the gas in pancakes, and the effects on the absorption profile,
can still be obtained from such simple approximations.

\section{ CONCLUSIONS}

In this paper, mini pancakes are analytically modeled with 
single plane wave perturbations, and the consequences for the 
$\lyal$ absorption
lines from these mini pancakes are discussed. Although a very crude
approximation to the real pancakes, this model is shown to elucidate
a number of important points.
The most important  results of the paper are contained in Figs. 3 and
5. They show
the mass scales of pancakes that corresponds to the HI column
densities of $\lyal$ forest lines with $\nha \la 10^{14.5}$ cm$^{-2}$,
the evolution of $\nha$ in time for different pancakes, and the
importance of bulk motions in low equivalent width absorption lines. 
It is shown that pancakes with
total masses in the range of $M \sim 10^{11} \hbox{--} 10{12.}$
M$_{\odot}$ (and a baryonic mass of $M_b \sim 10^{9.5} \hbox{--} 10^{10.5}$ 
M$_{\odot}$ ) can give rise to $\lyal$ lines with $\nha \la 10^{14.5}$
cm$^{-2}$. Here, the mass corresponds to the size of the perturbation.
These structures have an overdensity of the
order of $\sim 10$. Also, there is a maximum value of $\nha$ for a given 
value of
$\Omega _I$. For $\j21=1$, a temperature of the IGM of $\bar T =
10^4$ K, and $\Omega_I \sim 0.03$, mini pancakes only give
rise to $\lyal$ lines with $\nha \la 10^{14.5}$ cm$^{-2}$.
These 
results hold only for $z_c \sim 3$ for which the gas in the mini pancake
can be assumed to be compressed adiabatically.

Calculating the expected absorption profiles from the pancakes, we
also confirm that the information of the bulk motion associated with
pancakes is carried by low HI column density lines with
$\nha \la 10^{13}$ cm$^{-2}$. 
The lines in general consist of a central dip
with wings that signify bulk motions. The wings are most significant
for low equivalent width lines.
The lines could be fitted with more than one
Voigt profiles, but would not necessarily mean that the lines
arise in separate, discrete objects.

I am grateful to my colleagues in IUCAA and Dr. K. Subramanian
for valuable discussions, encouragement and criticisms. I also thank
the anonymous referee whose comments have improved the
manuscript. I am indebted to Prof.
P. J. E. Peebles for an inspiring discussion on $\lyal$ lines and pancakes.

\clearpage

\noindent
{\bf Figure captions}
\bigskip
\noindent
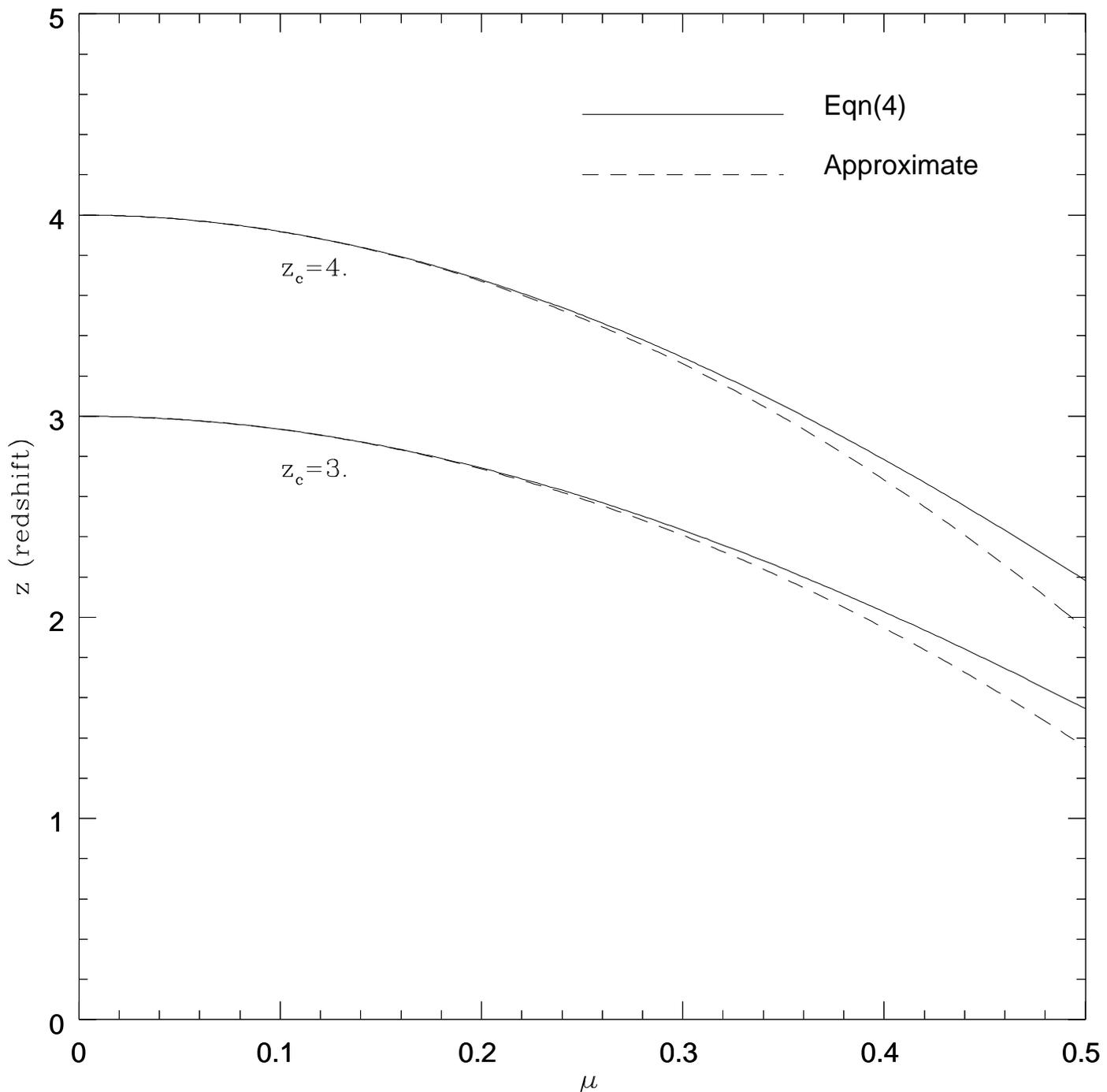
Fig. 1 -- The parameter $\mu$ is plotted against redshift $z$ for
two different collapse redshifts $z_c=3,4$. Solid curve is the result
of eqn ($4$) and the dashed curve corresponds to the approximation
used in the work.

\bigskip
\noindent
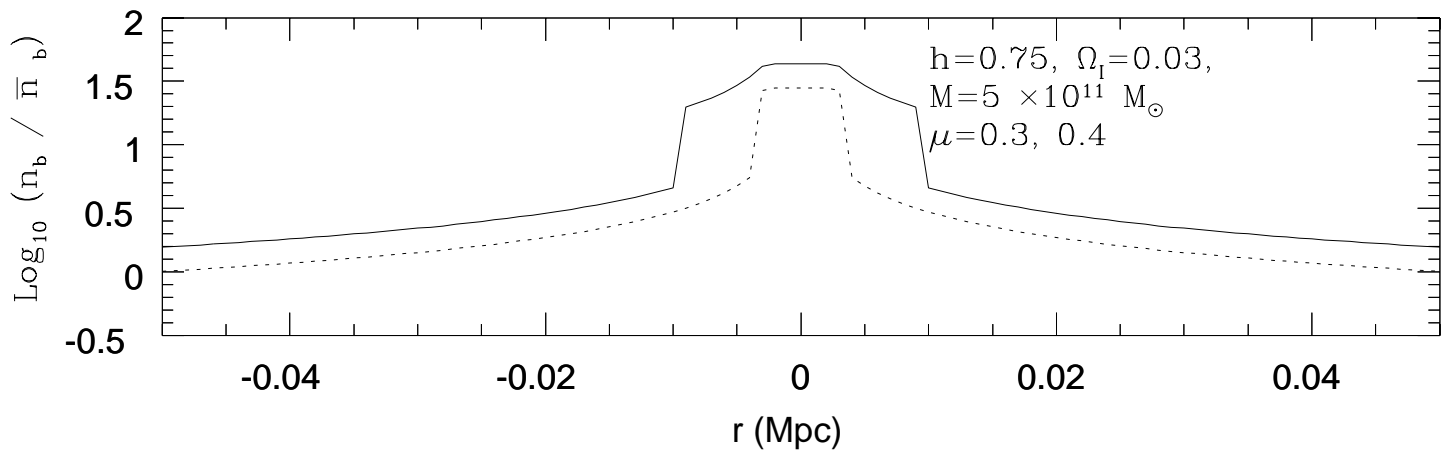
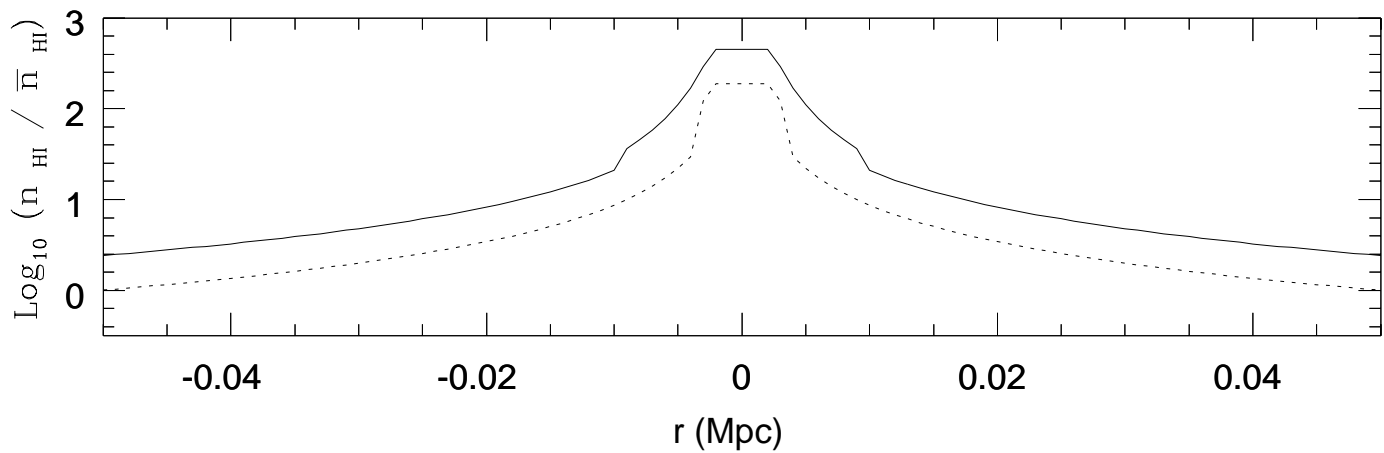
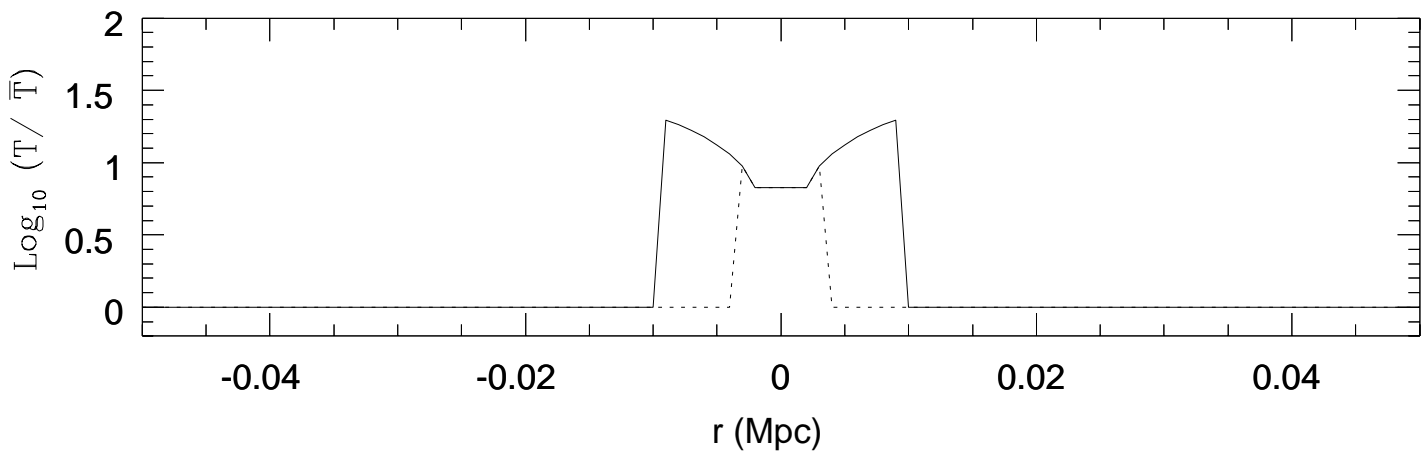
Fig. 2 -- (a) Curves show the overdensity profile of pancakes with
$M=5 \times 10^{11}$ M$_{\odot}$ at $\mu=0.3$ (solid curve)
and at $\mu=0.4$ (dotted curve). Most of the increase in the overdensity
in time comes from the decrease in the ambient density.

\medskip
\noindent
(b) Similar curves for H I overdensity profile.

\medskip\noindent
(c) Curves for the temperature profile for the same parameters. The
temperature of the IGM gas, $\bar T=10^4$ K.

\bigskip
\noindent
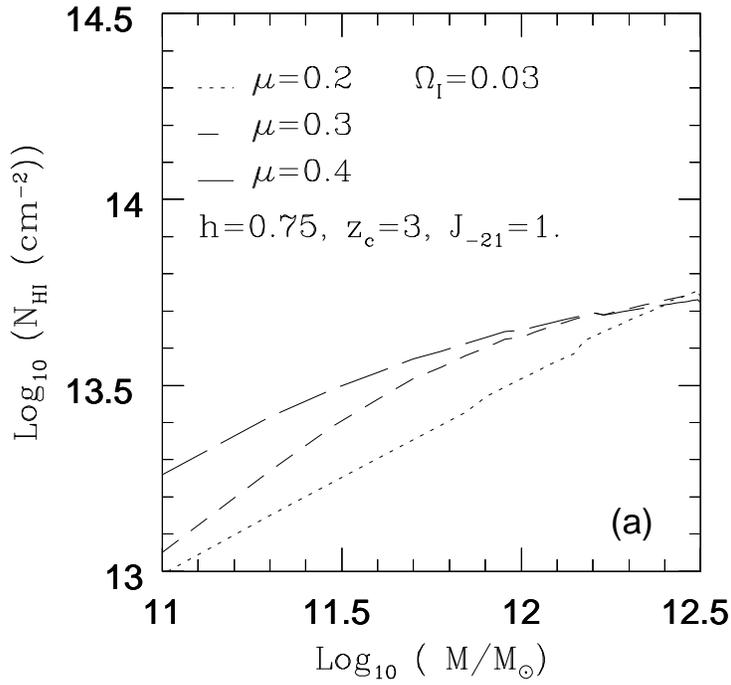
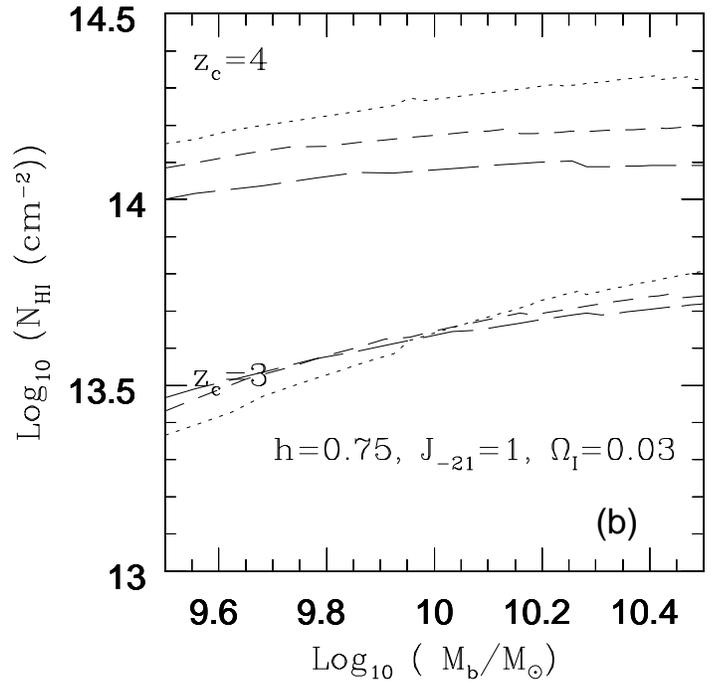
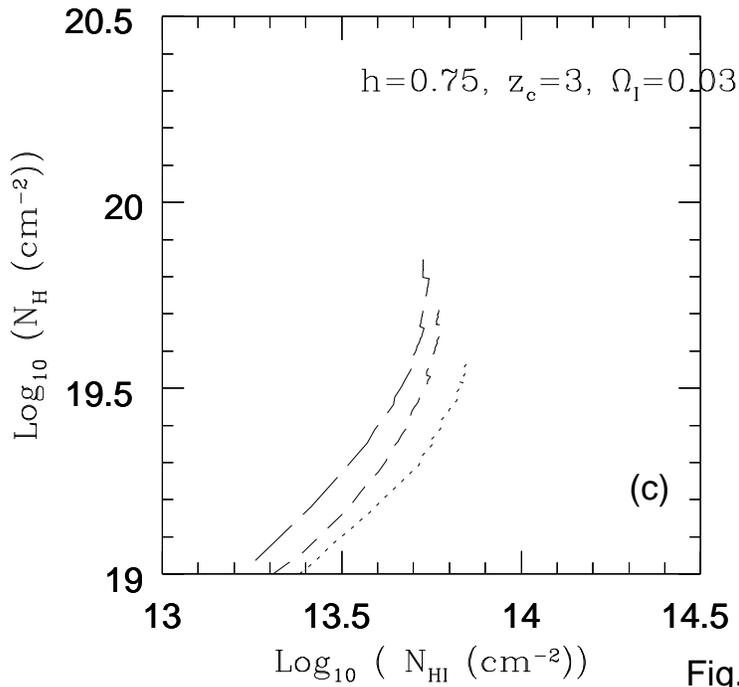
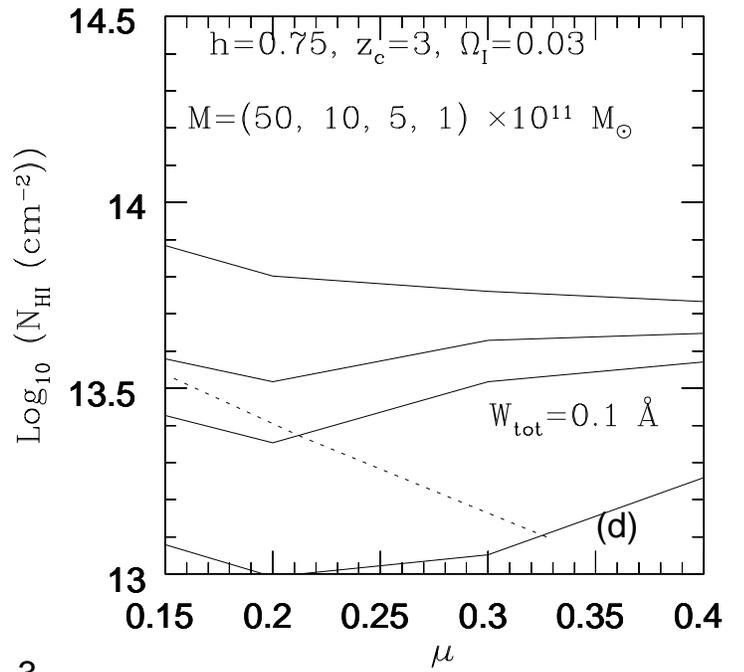
Fig. 3 -- (a) The H I column densities $\nha$ of pancakes, for regions
where the H I overdensity is greater than unity, are plotted against the
total mass $M$ of the pancakes for lines of sight perpendicular to 
the pancake. Dotted, short dashed and long dashed curves refer to pancakes 
with
$\mu=0.2, 0.3, 0.4$ respectively, for $\Omega _I=0.03$. The collapse
redshift of the pancakes $z_c=3$.

\medskip
\noindent
(b) Same as in Fig. 3(a), except that $\nha$ is plotted against the
baryonic masses $M_b= \mu \,\Omega_ I \, M$. The upper set of curves
correspond to collapse redshift $z_c=4$ and the lower set, to $z_c=3$.

\medskip \noindent
(c) Total column density $N_H$ is plotted against $\nha$ for $\mu=0.2, 0.3,
0.4$, $Omega_ I=0.03$, $z_c=3$.

\medskip \noindent
(d) The H I column densities $\nha$ of different masses are plotted
as functions of $\mu$ (or, equivalently, time). From top to bottom, the 
masses involved are $M=(50, 10, 5, 1) \times 10^{11}$ M$_{\odot}$.
The dotted curve
shows the contour left of which which the total equivalent width 
$W_{tot}$ of the 
corresponding $\lyal$ absorption profile is less than $0.1$ $\AA$. 

\bigskip 
\noindent
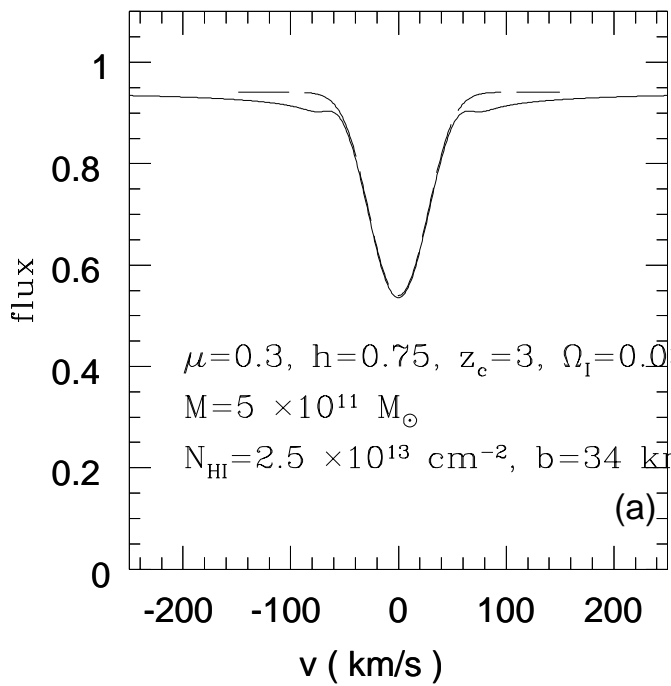
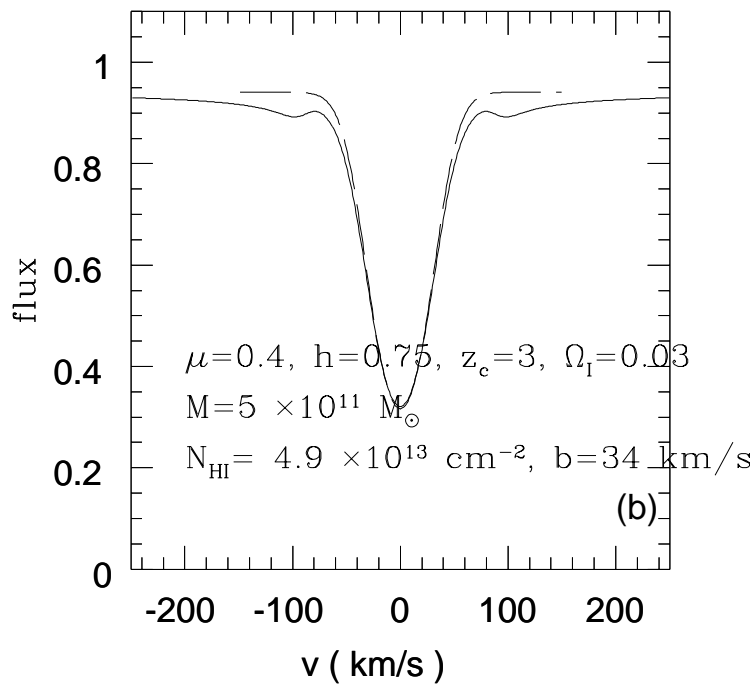
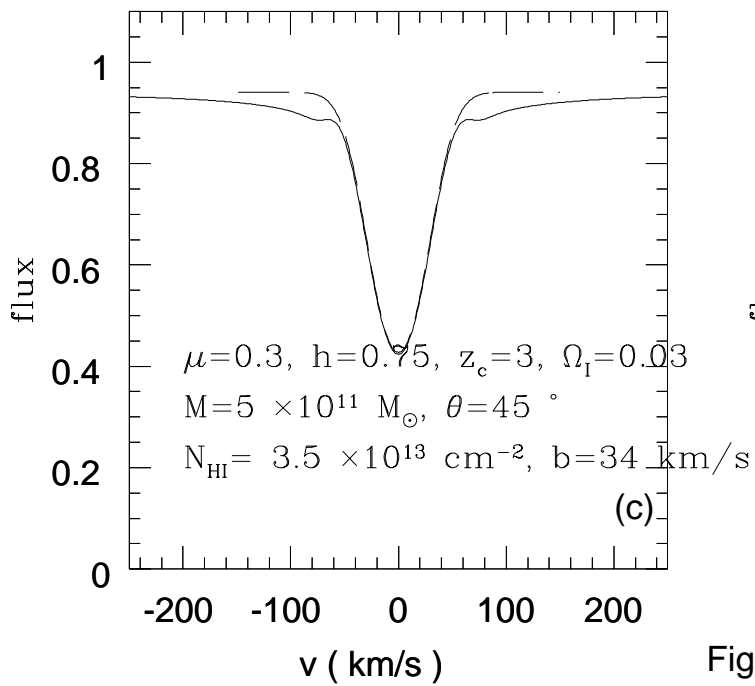
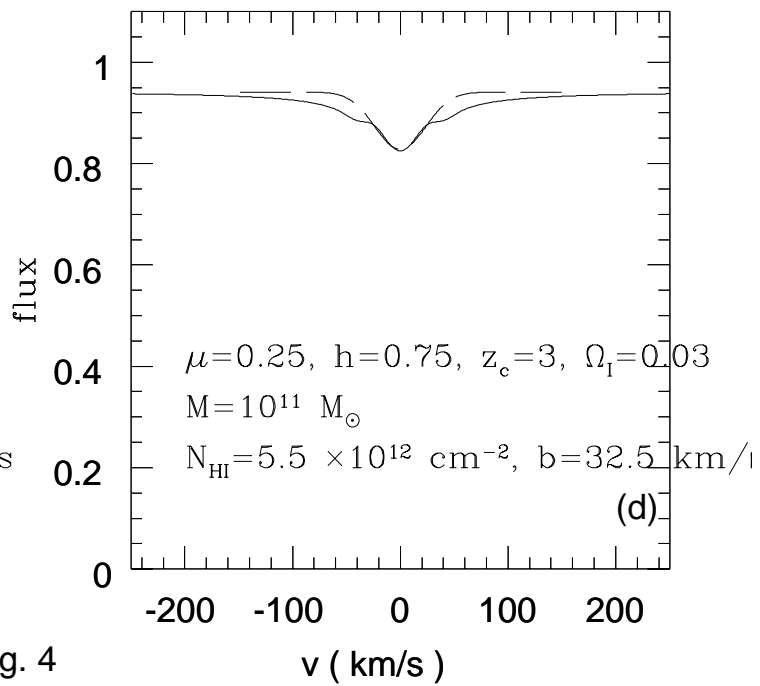
Fig. 4 -- (a) The $\lyal$ absorption profile for a pancake with
$M=5 \times 10^{11}$ M$_{\odot}$, $\mu=0.3$ for a line of sight
perpendicular to the pancake, for $z_c=3.$,
$h=0.75$, $\Omega_I=0.03$. The Voigt profile fitted to the central
dip has $\nha=2.4 \times 10^{13}$ cm$^{-2}$ and $b=35$ km/s.

\medskip \noindent
(b) The profile for the same pancake  but with $\mu=0.4$. The
Voigt profile has $\nha =10^{14}$  and $b=35$ km/s.

\medskip
\noindent
(c) The profile for the same pancake with $\mu=0.3$ but for
a line of sight making an angle $45 ^{\circ}$ to the pancake.
The Voigt profile fitted has $\nha=3.4 \times 10^{13}$ cm$^{-2}$
and $b \sim 35$ km/s.

\medskip \noindent
(d) The profile for a pancake with $M=10^{11}$ cm$^{-2}$
at $\mu=0.25$ (the cosmological parameters remain the same). The
Voigt profile fitted to the central dip has $\nha=3.6 \times
10^{12}$ cm$^{-2}$ and $b=30$ km/s.

\bigskip
\noindent
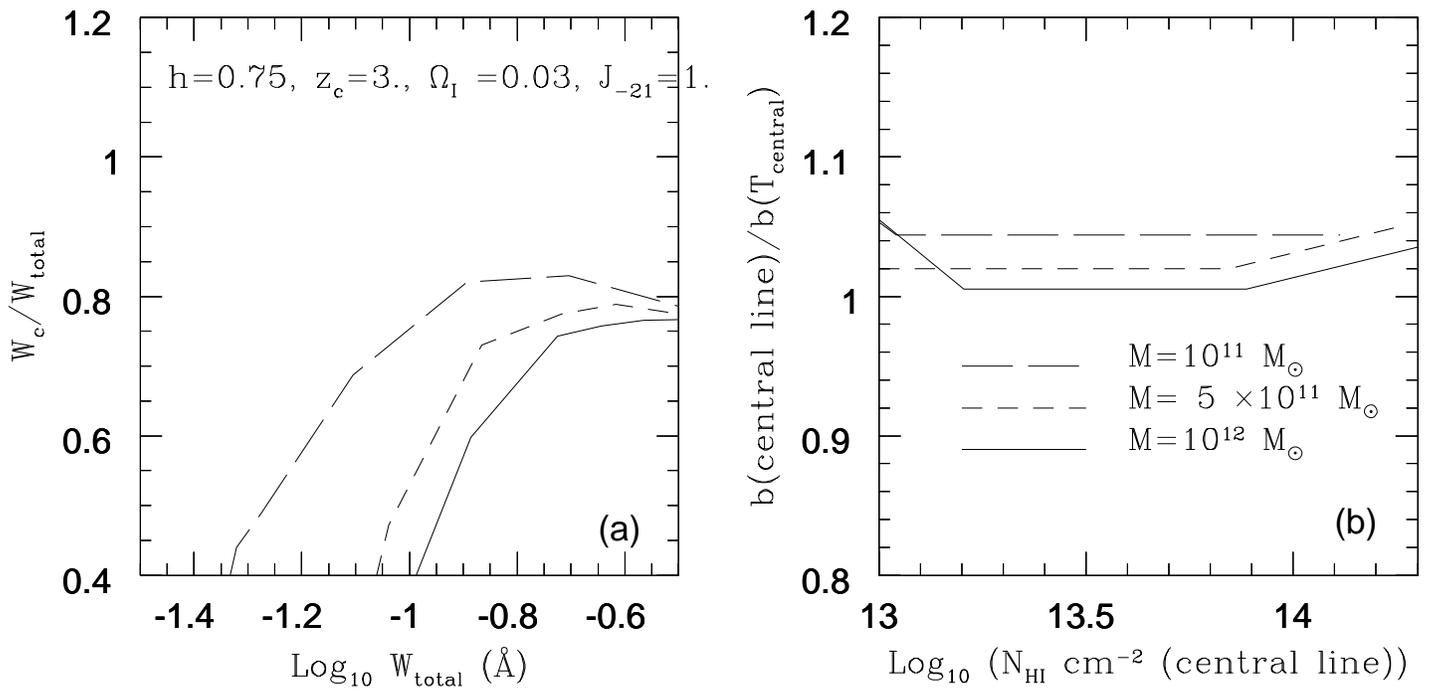
Fig. 5 -- (a) The ratio of the equivalent width of the fit to
the central dip $W_c$ to the total equivalent width of the absorption
profile $W_{tot}$ is plotted against $W_{tot}$ for different
masses. Curves with long dash, short dash and solid line correspond
to $M=10^{11}, 5 \times 10^{11}, 10^{12}$ M$_{\odot}$ with $
h=0.75$, $z_c=3$ and $\Omega _I=0.03$. 

\medskip
\noindent
(b) The ratio of the $b$ value of the Voigt fit to the central line
to the $b$ corresponding to $T_{max}$ of the pancake is plotted
against the values of corresponding $\nha$ of the central line.
The cosmological parameters are the same as in Fig. 5(a).

\medskip
\noindent
Fig. B -- The heating and cooling rate of gas with density $n_H=
10^{-4}$ cm$^{-3}$ in the presence of a photoionizing background
of intensity $\j21=1$ are plotted as functions of the temperature. 
The upper solid curve is the cooling
rate for $\j21=0$. The solid line at bottom is the cooling rate for
$\j21=1$ and the dashed line is the heating rate for the same ionizing
background.

\clearpage
\begin{appendix}

\section{APPENDIX A}

The derivations of eqns ($3$), ($4$), and ($5$) are sketched briefly
below, following ZS72 (see also Zeldovich and Novikov 1983). The
perturbation leading to the formation of the pancake is assumed to
be a sinusoidal plane wave in an $\Omega=1$ universe, with baryon
fraction $\Omega_b$. The motion of particles in the direction
of contraction (before being compressed
by shock waves) is given by
\begin{equation}
r=t^{2/3} \xi - b t^{4/3} \sin k \xi \> . 
\end{equation}
Here $\xi$ is the Lagrangian coordinate. The mass $M$ that is contained
in the region (of comoving length scale $\lambda$) of perturbation 
is given by,
\begin{equation}
M=\bar \rho \Bigl ( {\lambda \over 2} \Bigr ) ^3= \rho _o \Bigl ( {
\lambda (1+z) \over 2 } \Bigr )^3 \>; \bar \rho={1 \over
6 \pi G t^2} \>; \lambda = {2 \pi t^{2/3} \over k} \>. 
\end{equation}
Here, $\rho _o = 1.89 \times 10^{29} \> h^2 \> \Omega _I$ is 
the present day mass density of the IGM.
For matter not yet compressed by the wave, the density is given by
\begin{equation}
\rho={1 \over 6 \pi G t^2} \> {1 \over 1 - b t^{2/3} k \cos k \xi} \> .
\end{equation}
Therefore, $t_c$, the epoch of the formation of the pancake at $r=0$ is 
determined by $k b t_c ^{2/3}=1$. The parameter $\mu$ is then defined
as $\mu={k \xi \over \pi}$. This means that the interval $0 < \xi <
\pi /k$ in the Lagrangian coordinate is equivalent to $0 < \mu < 1$.

If the pressure of the IGM gas is assumed to be zero, then a plane
at $r=0$ reaches infinite density. One can think of a wave front in the
Lagrangian coordinate corresponding to the particles reaching this plane.
This wave front moves outward, and the Lagrangian coordinate $\mu _1$ of the
wave front is
given by,
\begin{equation}
r=0=\pi \mu _1 - (b k t_c ^{2/3}) \> \Bigl ( { t \over t_c } \Bigr )
^{2/3} \> \sin \pi \mu _1 \>. 
\end{equation}
Note that, in reality the shock will form at a distance from the $r=0$
plane, and the above expression, which assumes an origin at $r=0$ is
reasonable only for small gas pressure. However, as explained in the
text, we neglect gas pressure in the present work.
This shows that,
\begin{equation}
\Bigl ( { t \over t_c } \Bigr ) ^{2/3} = \Bigl ( { \pi \mu _1 \over
\sin \pi \mu _1 } \Bigr ) \>.
\end{equation}
Using $(A3)$, the density at the front is then given by,
\begin{eqnarray}
\rho _1 &= &{ \bar \rho \over 1- ({t \over t_c})^{2/3} \cos \pi \mu _1}
\nonumber\\
&=&{ \bar \rho \over 1- {\pi \mu _1 \over \tan \pi \mu _1}} \nonumber\\
&=& \rho _o { (1+z)^3 \> ( \sin \pi \mu _1 / \pi \mu _1 )^3 \over
(1- \pi \mu _1 / \tan \pi \mu _1)} \nonumber\\
&\sim& 3 \rho _o { (1+z)^3 \over \pi ^2 \mu _1 ^2} \>, \qquad\quad
\mu _1 \ll 1 \>. 
\end{eqnarray}

The velocity of matter running into the wavefront
 is obtained as, (for small $k \xi$ and
for $t \sim t_c$, {\it i.e.}, $\mu \ll 1$)
as,
\begin{eqnarray}
U_1= -{dr \over dt} &\approx& {2 \over 3} {\xi \over t^{1/3}} -{4 \over 3}
b t^{1/3} k \xi \nonumber\\
&=&{2 \over 3} {\xi \over t^{1/3}}-{4 \over 3} \Bigl ({t \over t_c} \Bigr )
^{1/3} {\xi \over t_c^{1/3}} \nonumber\\
&\sim& {2 \over 3} \Bigl ({t \over t_c} \Bigr )
^{1/3} {\xi \over t^{1/3}} \nonumber\\
&=&{2 \over 3} {1 \over t} { t^{2/3} \over k} \> (k \xi) \nonumber\\
&\approx &{2 \over 3 \> t} {\lambda \over 2 \pi} \Bigl (
{t \over t_c } \Bigr ) ^{1/3} \> ( \pi \mu _1) ^{1/2} \>  ( \sin \pi 
\mu _1)^{1/2} \nonumber\\
&=&23.3 \> h^{1/3} \> (1+z_c) ^{1/2} \> ( \pi \mu _1) ^{1/2} \>  ( \sin \pi 
\mu _1)^{1/2} \> \Bigl ( { M \over 10^{11} \> M _{\odot} } \Bigr )^{1/3}
\> {\rm km/s} \>. 
\end{eqnarray}
Here $(1+z_c)$ corresponds to $t_c$.

For matter with non-zero pressure, the above equations will remain
approximately unchanged for large $\mu$, when the wave has moved a large
distance. The central regions will, however, not attain infinite density
in this case (see \S 2.3).

\bigskip
\section{APPENDIX B}

We present the cooling curve for a gas in the presence of ionizing UV
photons here. The cooling processes used in the calculation include
collisional ionization, recombination, dielectronic recombination and
collisional excitation, in a gas containing hydrogen and helium 
($X=0.76$), as listed in Black (1981). We used the photoionization
cross-section for $\j21=1$ as in Black (1981) (for a spectral index
$\alpha=1$). Figure B shows the cooling rate (solid lines) for a gas with 
$n_H =10^{-4}$ cm${-3}$ with $\j21=0$ and $\j21=1$, and shows that
line cooling is suppressed by the ionizing photons substantially. For higher
densities, the suppression of line cooling is less. Therefore, for a maximum
density of $\sim 10^{-4}$ cm$^{-3}$ and temperature $\la 10^6$ K, 
as in eqn ($8$), the cooling rate of the gas is $\sim n_H^2 10^{-24}
\sim 10^{-32}$ erg cm$^{-3}$ s$^{-1}$. This implies a cooling time scale
of $10^{17} \> (T/ 10^5 \> k)$ s, which is larger than the Hubble
time at $z \sim 3$.

Fig. B also plots the heating rate (dashed line) for the same density
and for $\j21=1$. It is seen that at $T \sim 10^{4.7}$ and $n=10^{-4}$
cm$^{-3}$ (eqn ($8$), the heating
time scale is ${3/2 \, n k T \over \Gamma} \sim
3 \times 10^{16}$ s, comparable to the Hubble
time at $z \sim 3$ for a flat universe and $h=0.75$. 

Therefore, the assumption of adiabaticity for the gas in mini pancakes
is reasonable only for $z_c \sim 3$.

\end{appendix}

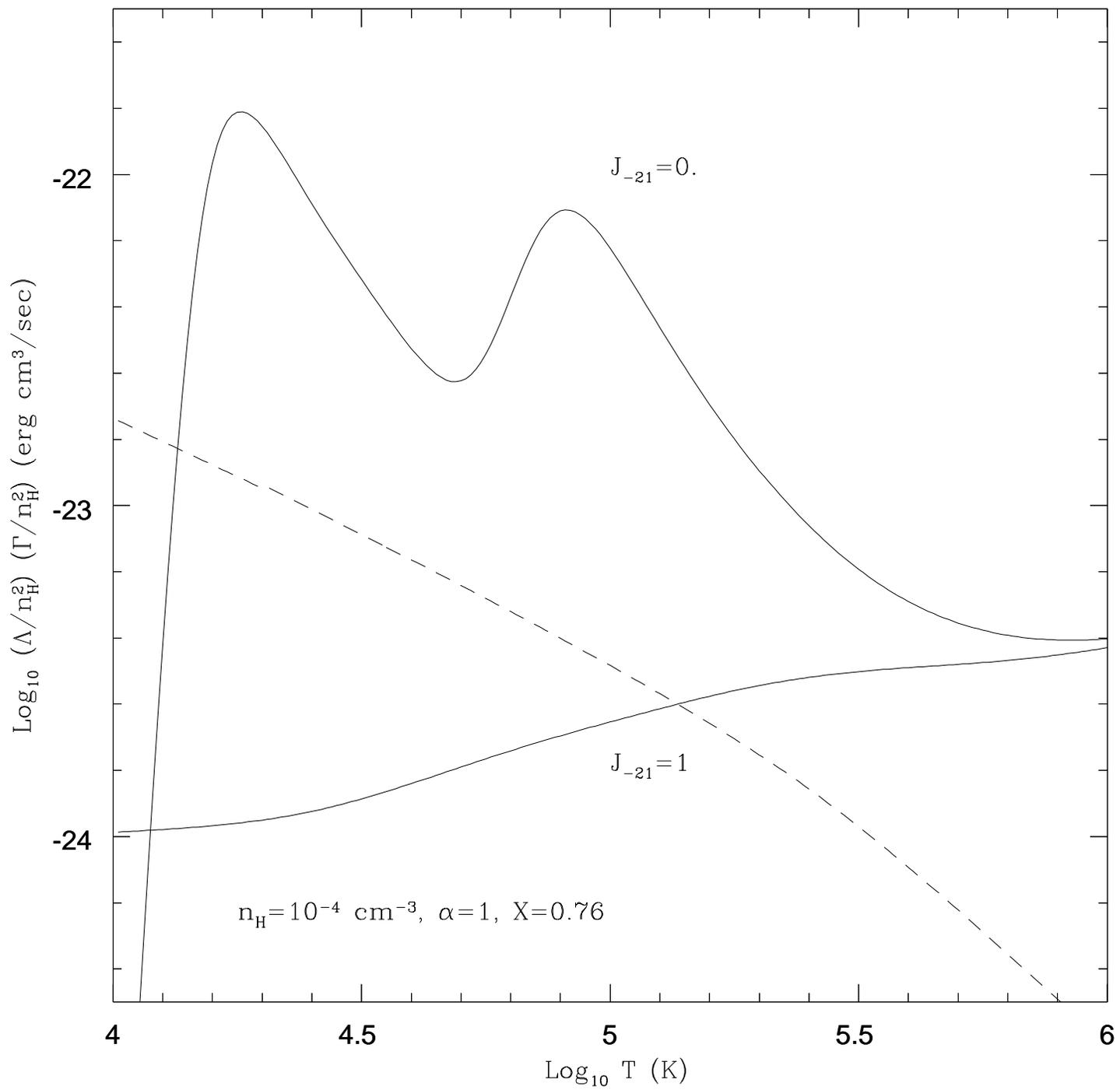
\end{document}